\documentclass[letterpaper, 10 pt, conference]{ieeeconf}  

\IEEEoverridecommandlockouts                              

\usepackage{amsmath} 
\usepackage{amssymb}  
\usepackage{mathtools}
\usepackage{graphicx}
\usepackage{color}
\usepackage{xcolor} 
\usepackage{cite}
\usepackage{acronym}
\usepackage{subcaption}
\usepackage{siunitx}
\usepackage{tikz}
\usetikzlibrary{
    arrows,
    arrows.meta,
    calc,
    patterns,
}
\usepackage{pgfplots}
\pgfplotsset{compat=newest}

\acrodef{IMU}{inertial measurement unit}
\acrodef{e-scooter}{electric scooter}

\title{\LARGE \bf On the effects of angular acceleration in orientation estimation using inertial measurement units
}

\author{Felix Brändle, David Meister, Marc Seidel, Robin Strässer, Frank Allgöwer
\thanks{F.\ Allgöwer is thankful that this work was funded by the Ministry of Science, Research and the Arts of the State of Baden-Württemberg (MWK) in the context of the ``MobiLab'' Project.
F.\ Brändle thanks the International Max Planck Research School for Intelligent Systems (IMPRS-IS) for its support.
D.\ Meister, M.\ Seidel, R.\ Strässer thank the Graduate Academy of the SC SimTech for its support.}
\thanks{F.\ Brändle, D.\ Meister, M.\ Seidel, R.\ Strässer, and F.\ Allgöwer are with the University of Stuttgart, Institute for Systems Theory and Automatic Control, 70550 Stuttgart, Germany
(e-mail: e-scooter@ist.uni-stuttgart.de).}%
}

\begin{document}
 \pubid{\begin{minipage}{\textwidth}\ \\[60pt] \copyright 2025 IEEE. This version has been accepted for publication in 9th IEEE Conference on Control Technology and Applications (CCTA) 2025. Personal use of this material is permitted. Permission from IEEE must be obtained for all other uses, in any current or future media, including reprinting/republishing this material for advertising or promotional purposes, creating new collective works, for resale or redistribution to servers or lists, or reuse of any copyrighted component of this work in other works.\end{minipage}}	
	
\maketitle

\begin{abstract}
    In this paper, we analyze the orientation estimation problem using \acl{IMU}s.
    Many estimation algorithms suffer degraded performance when accelerations other than gravity affect the accelerometer.
   	We show that linear accelerations resulting from rotational accelerations cannot be treated as external disturbance to be attenuated, rather, they change the dynamic behavior of the filter itself.
    In particular, this results in the introduction of additional zeros in the linearized transfer functions.
    These zeros lead to non-minimum phase behavior, which is known to be challenging for control.
    We validate these findings experimentally.
    Further, we demonstrate that Mahony and Madgwick filters can attenuate the acceleration at the expense of reduced bandwidth.
    In addition, we show that validation schemes based on pre-collected data fail to capture these closed-loop effects accurately.
\end{abstract}


\section{INTRODUCTION}
Determining the rotation of a rigid body is a common problem in engineering and finds application in robotics, unmanned vehicles, human motion tracking, and quadcopters~\cite{Ludwig2018a, Nazarahari2021}.
One approach to estimate the orientation is to use an \ac{IMU} consisting of an accelerometer and a gyroscope.
If available, an additional magnetometer can be added to improve accuracy and to estimate the yaw.
By applying a sensor fusion algorithms such as the Mahony filter~\cite{Mahony2008}, the Madgwick filter~\cite{Madgwick2011}, or a Kalman filter~\cite{Ludwig2018a}, these measurements can be combined to obtain an accurate estimate of the body's true orientation.
Integrating the angular velocity from the gyroscope is inaccurate due to drift.
Hence, the accelerometer is used to account for the error.
This approach is based on the assumption, that the \ac{IMU} is at rest, meaning the only acceleration acting on the \ac{IMU} is gravity.
Since the gravitation vector is known, it is possible to determine the angular displacement of the sensor to a reference frame.

A known challenge arises if additional accelerations beyond gravity affect the sensor.
Since the at-rest condition no longer holds, the filter performance deteriorates.
Several methods have been proposed to address this, such as optimizing over the filter parameters~\cite{Ludwig2018}, compensating for additional accelerations using a model~\cite{Ahmed2017, Briales2021}, or using  an adaptive schemes~\cite{Wei2025, Candan2021, Park2020}.
Model-based schemes require a model of the system and therefore must be designed for the specific application.
Hence, off-the-shelf filters cannot be used.
On the other hand, adaptive schemes treat the acceleration as an external disturbance independent of the quantities to be estimated.
When the filter is used in a closed loop together with a controller, an independent disturbance cannot destabilize the system, at least for linear systems and, by extension, nonlinear systems close to an operating point.
Yet, when the accelerometer is not located at the rotational axis, any angular acceleration results in a linear acceleration depending on the angles to be estimated, hence it is not an independent disturbance.
This leads to a different dynamical system with different properties, which may cause instabilities when combined with a controller.

One such example of angular accelerations is the Cubli~\cite{Gajamohan2012}.
Due to physical limitations, it is not feasible to place the \ac{IMU} exactly on the rotational axis, leading to linear accelerations induced by rotational motions.
Interestingly, these accelerations can be compensated for by employing multiple \acp{IMU}~\cite{Gajamohan2012, Trimpe2010}.
However, there is no analysis of the effects of why this compensation is indeed necessary, besides the need to compensate for all accelerations except the gravitational one.

In this work, we investigate the effects of angular accelerations when the \ac{IMU} is not positioned at the rotational axis.
As our main contribution, we show that the angular accelerations lead to a qualitative change of the estimators behavior by adding additional unstable zeros to the transfer function.
We further demonstrate that this change negatively affects feedback systems in particular.
In addition, we analyze how the filter parameters can help mitigating these undesirable effects and outline what trade-off must be made to suppress the effects of the angular acceleration.
In particular, our analysis offers insights and tuning guidelines for the Mahony and Madgwick filters.
Then, we verify the presented results on a real system, where we recover the discussed behavior.
We expect those insights to be transferable to a wide range of practical systems, leading to more robust and more accurate \ac{IMU}-based estimator.

\emph{Notation:}
We denote the $n\times m$ zero matrix as $0_{n\times m}$, where we omit the indices if the dimensions are clear from context.
Moreover, we use $q\in\mathbb{R}^4$ to denote a unit quaternion and $\hat{q}\in\mathbb{R}^4$ for its estimate. 
In addition, $\otimes$ denotes quaternion multiplication and $q^\star$ the conjugate of a quaternion $q$, cf.\ \cite{Berner2007}.
Further, $a \times b$ is the vector product of $a\in\mathbb{R}^3$ and $b\in\mathbb{R}^3$.
We write $a_{i:j}$ for the $i$th to $j$th component of the vector $a\in\mathbb{R}^n$, where $1\leq i<j\leq n$. 
Finally, $\alpha = \mathrm{atan2}(y,x)$ is the angle $\alpha$ between the positive $x$-axis and the connection from the origin to the point $(x,y)$, where $-\pi<\alpha\leq \pi$.
\section{MODEL}\label{sec:Model}
\begin{figure}[t]
	\centering
	\begin{tikzpicture}[scale=0.9]
	
	\def\varphiang{-30}
	\def\height{4}
	\def\rArc{0.5*\height}
	
	\def\groundlength{1}
	\def\groundheight_plot{0.12}
	
	\def\supportLength{0.4}
	
	\coordinate (origin) at (0,0);
	\coordinate (originKOS) at ($(origin)$); 
	\coordinate (IMU) at ($(origin) + ({\height*sin(\varphiang)},{\height*cos(\varphiang)})$);
	
	\draw[thick] (origin) -- (IMU) node[midway,left] {$l$};
	
	\draw[dashed](origin) --++ ($({0},{\height*cos(\varphiang)})$);
	
	\draw[-Stealth] ($(origin) + ({0},{\rArc})$) arc(90:90-\varphiang:\rArc) node[midway, above]{$\varphi$};
	
	\node[draw,fill = white, rotate = -\varphiang] at (IMU) {IMU};
	
	\draw[-Stealth,thick] (originKOS) --++ (1,0) node[below]{$e_y^I$};
	\draw[-Stealth,thick] (originKOS) --++ (0,1) node[right]{$e_z^I$};
	
	\draw[-Stealth,thick] (originKOS) --++ ($({cos(\varphiang)},{-sin(\varphiang)})$) node[above]{$e_y^K$};
	\draw[-Stealth,thick] (originKOS) --++ ($({sin(\varphiang)},{cos(\varphiang)})$) node[left]{$e_z^K$};
	
	\fill (origin) circle(1.5pt) node[yshift = -2, xshift = -2, above left]{$O$};
	

	\coordinate (groundStart) at ($(origin)-(0,{\supportLength*sin(60)})-({0.5*\groundlength},0)$);	
	
	\fill[pattern=north east lines] (groundStart) rectangle ++(\groundlength,-\groundheight_plot);	
	\draw[thick] (groundStart) -- ++(\groundlength,0);	
	
	\draw[thick] (origin) --++ ($({\supportLength*cos(60)}, {-\supportLength*sin(60)})$) --++ ($(-\supportLength,0)$) -- (origin);
	
	\draw[thick,-stealth] ($(origin)+({0.5*\groundlength},{\height*cos(\varphiang)})$) --++(0,-1) node[midway, left]{$\mathrm{g}$};
	
\end{tikzpicture}
	\caption{Pendulum with \acs{IMU} including gravity $\mathrm{g}$.}
	\label{IMG:Setup:Pendulum}
\end{figure}
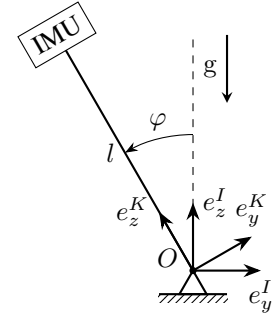

To model the effects of angular motion around a fixed axis on the estimation, we consider a pendulum system, as illustrated in Fig.~\ref{IMG:Setup:Pendulum}.
This setup serves as a generalized model for various types of rotational motion that are not centered at the rotational axis, e.g., loops in aerobatics, joint motions in exoskeletons, or curve driving in vehicles \cite{Nazarahari2021, Gajamohan2012}, such that our results are transferable to different systems.
The \ac{IMU} is placed at a fixed distance $l$ to the point $O$ and rotates with respect to the roll angle $\varphi$ while pitch $\theta$ and yaw $\psi$ are held at zero.
Hence, any acceleration in $\varphi$ causes a linear acceleration that is measured by the accelerometer.

We align the pendulum's rotation with the roll angle to investigate whether an acceleration in $\varphi$ influences the estimated pitch and yaw.
Note that a similar analysis can be performed around any axis, with the exception of the yaw, which cannot be estimated using accelerometers alone.
Reconstructing the yaw angle requires additional sensors, e.g., a magnetometer, which also provides redundancy the pitch estimation~\cite{Madgwick2011}. 
An investigation including magnetometers is beyond the scope of this paper and is left for future research.
The \ac{IMU} records both the linear acceleration $a^K$ including gravity and the angular velocity $\omega^K$, expressed in the body-fixed coordinate system $K$, which is rotated relative to the inertial frame $I$. 
The measurement equations are given by
\begin{equation}
	a^K = \begin{pmatrix}
		0 \\
		\sin(\varphi) - \frac{l}{g}\ddot{\varphi} \\
		\cos(\varphi) - \frac{l}{g}\dot{\varphi}^2
	\end{pmatrix} 
	,\qquad 
	\omega^K = \begin{pmatrix}
		\dot{\varphi}\\ 
		0 \\
		0
	\end{pmatrix}.
\end{equation}
The acceleration measurements are already expressed in units of standard gravity [$g$].
We assume that $\varphi$ is twice differentiable, ensuring the existence of $\ddot{\varphi}$.
Since the acceleration depends only on the ratio between $l$ and $g$, we normalize $g$ to $1$ for simplicity throughout the paper. 
A common assumption in literature is that the \ac{IMU} is at rest, meaning $\dot{\varphi}=\ddot{\varphi}=0$ such that gravity is the only source of acceleration affecting $a^K$~\cite{Madgwick2011, Mahony2008, Ludwig2018a}.
Under this condition, it is possible to compute $\varphi$ using simple trigonometric relations on $a^K$.
In the remainder of the paper, we analyze the effects when the \ac{IMU} is not at rest.

\section{FILTER ANALYSIS} \label{sec:Analysis}
Next, we analyze filters using a quaternion-based formulation. 
Following~\cite{Berner2007}, it is possible to transform quaternions back and forth between roll, pitch, and yaw angles. 
The corresponding relationships are
\begin{subequations}
	\begin{align}
		\varphi &= \mathrm{atan2}\left(q_1 q_2 + q_3 q_4,\tfrac{1}{2}-q_2 ^2 - q_3 ^2\right), \label{eq:analysis:phi}\\
		\theta &= \mathrm{arcsin}\left(2(q_1 q_3 - q_2 q_4)\right), \\
		\psi &= \mathrm{atan2}\left(q_1 q_4 + q_2 q_3 ,\tfrac{1}{2}-q_3 ^2 - q_4 ^2\right).
	\end{align}	
\end{subequations}
A common preprocessing step is to first normalize the acceleration measurement ~\cite{Mahony2008,Madgwick2011,Euston2008}
\begin{equation}
	\bar{a}^K = \frac{a^K}{\|a^K \|}.
\end{equation}

\subsection{One-dimensional case}\label{sec:Analysis:OneDim}
\begin{figure}[tb]	
	\centering	
	\includegraphics[scale = 0.44]{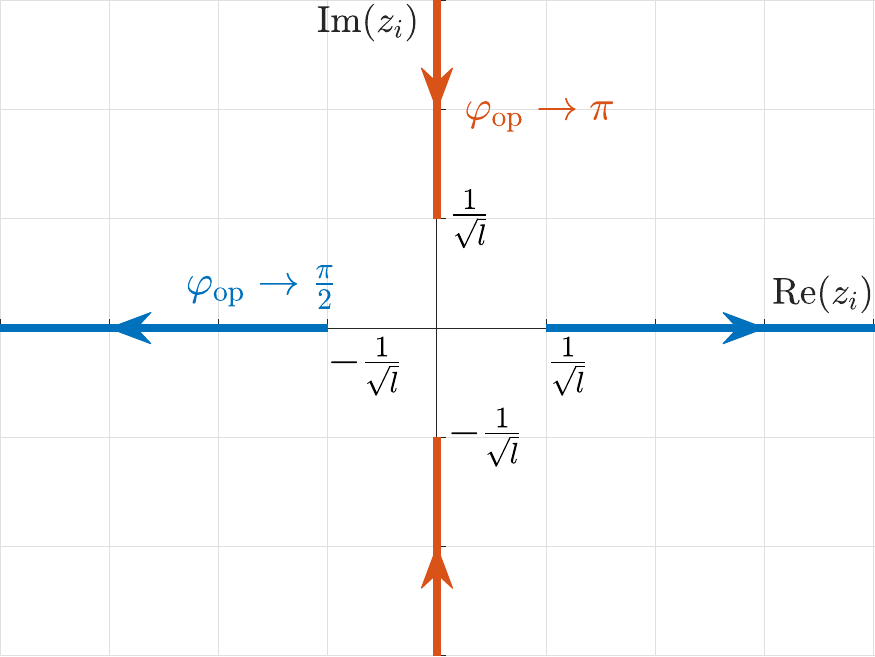} 
	\caption{Location of the zeros for different operating points $\varphi_{\mathrm{op}}\in [ 0,\pi ]$ using~\eqref{eq:Ana:tfAtan}.}
	\label{IMG:Ana:ZerosAtan}
\end{figure}

Before introducing quaternion based formulations for three-dimensional rotation, we consider the simpler one-dimensional case. 
Later, we show that in the full quaternion formulation, accelerations in the roll angle do not affect the estimate of the yaw and pitch.
In particular, we estimate the angle $\varphi$ solely from the normalized acceleration $\bar{a}^K$.
If the \ac{IMU} is at rest, i.e., $\dot{\varphi}=\ddot{\varphi}=0$, it is possible to recover $\varphi$ exactly as
\begin{equation}
	\varphi = \mathrm{atan2}(\bar{a}^K_2, \bar{a}^K_3).
\end{equation}
To investigate the effect of $\dot{\varphi}$ and $\ddot{\varphi}$, we linearize around an operating point $\varphi=\varphi_{\mathrm{op}}$ with $\dot{\varphi}_{\mathrm{op}}=\ddot{\varphi}_{\mathrm{op}}=0$ and apply the Laplace transform.
This yields the following frequency-domain behavior for the roll estimate $\hat{\varphi}$
\begin{equation}\label{eq:Ana:tfAtan}
	\Delta \hat{\varphi}(s) = (1-l\cos(\varphi_{\mathrm{op}})s^2) \Delta \varphi(s)	
\end{equation}
with $\Delta \hat{\varphi}=\hat{\varphi}-\varphi_\mathrm{op}$ and $\Delta \varphi = \varphi - \varphi_{\mathrm{op}}$.
The transfer function has zeros at $z_{1,2}=\pm\frac{1}{\sqrt{l\cos(\varphi_{\mathrm{op}})}}$.
Fig.~\ref{IMG:Ana:ZerosAtan} shows the location of the zeros for different operating points.
For $\varphi_{\mathrm{op}} = 0$, we obtain two zeros at $\pm \frac{1}{\sqrt{l}}$, representing one real unstable and one real asymptotically stable zero.
This leads to deviations from the ideal constant transfer function  $G_{\Delta \varphi \to \Delta \hat{\varphi}}(s) \equiv 1$.
At low frequencies, the zeros do not affect the estimate.
Since the zeros are symmetric about the imaginary axis, the phase change cancel each other out, resulting in only magnitude amplifications for high frequencies.
As $\varphi_{\mathrm{op}}\to\frac{\pi}{2}$, the zeros move further away from the origin, resulting in a transfer function closer to the ideal transfer function $1$ for the intermediate frequencies.
For $\varphi_{\mathrm{op}}=\frac{\pi}{2}$ the term $l\cos(\varphi_{\mathrm{op}})$ vanishes, and the transfer function simplifies to the ideal case $G_{\Delta \varphi \to \Delta \hat{\varphi}}(s) \equiv1$ completely unaffected by $\ddot{\varphi}$.

For the operating point $\varphi_{\mathrm{op}}\in (\frac{\pi}{2},\pi]$, the behavior of the system changes qualitatively.
This is because the $z$-axis of the \ac{IMU}'s body frame $K$ is no longer pointing upward but downward.
As a result, the zeros change from being real numbers to being purely imaginary and tend towards $\pm i \frac{1}{\sqrt{l}}$ as $\varphi_{\mathrm{op}}\to\pi$.
As with the upper position $\varphi_\mathrm{op}=0$ , the magnitude of high-frequency compenents is amplified due to $\ddot{\varphi}$.
However, there is a phase change of $180^\circ$ for frequencies larger than $\SI[parse-numbers = false]{\frac{1}{\sqrt{l}}}{\radian\per\second}$, i.e., the sign of the roll angle $\varphi$ and its estimate $\hat{\varphi}$ differ in sign and magnitude.
Further, sinusoids with a frequency $\SI[parse-numbers = false]{\frac{1}{\sqrt{l}}}{\radian\per\second}$ are perfectly canceled by the zero, resulting in $\hat{\varphi}=0$, even though $\varphi\neq 0$.
Note that the zeros scale with $\frac{1}{\sqrt{l}}$.
Thus, reducing $l$ also shifts the zeros to higher frequencies, improving filter accuracy for the intermediate frequencies.
This is intuitive, as placing the $\ac{IMU}$ closer to the rotation axis reduces the effect of $\ddot{\varphi}$.

After establishing the most basic estimation procedure, we extend our analysis to estimation of all angles and including the gyroscope.
To this end, we consider two common filters, the Mahony filter~\cite{Mahony2008} and the Madgwick filter~\cite{Madgwick2011}, and discuss how the corresponding tuning parameters can improve the rejection of angular accelerations.

\subsection{Mahony filter}
The Mahony filter is a particular filter structure based on proportional $k_p$ and integral $k_i$ gains, allowing the estimated quaternion $\hat{q} $ to track the orientation that aligns the gravity vector with the measured direction $\bar{a}^K$ using a PI-controller.
The resulting filter is described by the differential equation~\cite{Mahony2008}
\begin{equation}\label{eq:Mahony-ode}
	\begin{pmatrix}
		\dot{\hat{q}}  \\
		\dot{\zeta}
	\end{pmatrix} 
	= 
	\begin{pmatrix}
		\frac{1}{2} \hat{q} \otimes \begin{pmatrix}0 \\ \omega^K + \zeta\end{pmatrix} \\
		0_{3\times 1}
	\end{pmatrix}+
	\begin{pmatrix}
		\frac{1}{2} \hat{q} \otimes \begin{pmatrix}0 \\ k_p e\end{pmatrix} \\
		k_i e
	\end{pmatrix}
\end{equation}
with
\begin{equation}
	e = \bar{a}^K \times  \left(\hat{q} ^\star \otimes \begin{pmatrix}0_{3\times1} \\ 1 \end{pmatrix} \otimes \hat{q} \right)_{2:4}.
\end{equation}
This formulation can also be interpreted as a disturbance observer~\cite{Andrievsky2020}.
The first term of~\eqref{eq:Mahony-ode} represents the nominal prediction including a constant disturbance in $\omega^K$, while the second term acts as correction due to the observed measurement error $e$.

To analyze the estimator, we linearize the filter around $\varphi=\varphi_{\mathrm{op}}$ with $\dot{\varphi}_{\mathrm{op}}=\ddot\varphi_{\mathrm{op}}=0$ to obtain a linear differential equation for the roll, pitch, and yaw dynamics
\begin{equation}\label{eq:Ana:ODEMahony}
	\begin{aligned}
		\Delta\dot{\hat{x}} 	&= A\Delta \hat{x} + B \Delta \bar{a}^K, \\
		\Delta y 		&= C\Delta \hat{x}.		
	\end{aligned}
\end{equation}
where $\Delta\hat{x}$ contains the stacked linearized estimates $\Delta\hat{q}$ and $\Delta\hat{\zeta}$. 
Note that other estimation schemes, such as Luenberger observers and Kalman filters, exhibit the same linear form after linearization.
The gains $k_p$ and $k_i$ are just a particular choice of the observer gains.
For interpretability, we keep $k_p$ and $k_i$ in our analysis.
System~\eqref{eq:Ana:ODEMahony} can also be represented via its corresponding transfer function. 
The corresponding transfer function can be derived by substituting the linearized $\Delta \bar{a}^K$, resulting in
\newlength\mylenA
\settoheight\mylenA{$
	\begin{pmatrix}
		\Delta \hat{\varphi}(s) \\
		\Delta \hat{\theta}(s) \\
		\Delta \hat{\psi}(s)
	\end{pmatrix} 
	=
	\begin{pmatrix}
		\frac{k_i + k_ps + (1-k_il\cos(\varphi_{\mathrm{op}}))s^2 - k_pl\cos(\varphi_{\mathrm{op}})s^3}{k_i + k_ps + s^2} \\ 
		0 \\
		0 
	\end{pmatrix} \Delta \varphi(s).
$}
\begin{equation*}
\resizebox{\linewidth}{\mylenA}{$
	\begin{pmatrix}
		\Delta \hat{\varphi}(s) \\
		\Delta \hat{\theta}(s) \\
		\Delta \hat{\psi}(s)
	\end{pmatrix} 
	=
	\begin{pmatrix}
		\frac{k_i + k_ps + (1-k_il\cos(\varphi_{\mathrm{op}}))s^2 - k_pl\cos(\varphi_{\mathrm{op}})s^3}{k_i + k_ps + s^2} \\ 
		0 \\
		0 
	\end{pmatrix} \Delta \varphi(s).
$}
\end{equation*}
We observe that the pitch and yaw estimates are zero, indicating no coupling between $\varphi$ and the yaw, and pitch estimate.
Therefore, we restrict our analysis to the transfer function from $\Delta\varphi$ to $\Delta\hat{\varphi}$.
As expected, the poles depend on $k_p$ and $k_i$, which much be chosen to ensure asymptotic stability.
However, the zeros depend on $l$ and the chosen operating point, i.e., they can not be moved arbitrarily by the filter gains without changing the poles.
These zeros fundamentally shape the filter’s dynamic response to angular accelerations.
Interestingly, for $\varphi_{\mathrm{op}}=\frac{\pi}{2}$ the nominator and the denominator cancel perfectly, resulting in the ideal transfer function $G_{\Delta \varphi \to \Delta \hat{\varphi}}(s) \equiv 1$, which coincides with the results discovered for the one-dimensional case.
This means $\ddot{\varphi}$ does not affect the estimate around $\varphi_\mathrm{op}=\frac{\pi}{2}$.
Further, we set $k_i=0$, i.e., we omit the integrator.
This is reasonable since integral action primarily compensates for low-frequency disturbances, such as gyroscope bias, which we do not consider in this work.
This reduction leaves $k_p$ as the sole tuning parameter.

Exploiting the prior discussion, the zeros of the transfer function $G_{\Delta \varphi \to \Delta \hat{\varphi}}(s)$ are determined by
\begin{equation}
	k_p + s  - k_p l \cos(\varphi_{\mathrm{op}}) s^2 = 0,
\end{equation}
where $k_p>0$ ensures asymptotic stability.
The corresponding roots follow as
\begin{equation}
	z_{1,2} = \frac{-1\pm\sqrt{1+4k_p^2l\cos(\varphi_{\mathrm{op}})}}{2k_p l \cos(\varphi_{\mathrm{op}})}.
\end{equation}
As before, we consider two ranges for the operating point.
If $\varphi_{\mathrm{op}}\in[0,\frac{\pi}{2})$, both zeros are real-valued.
For small $k_p$, one zero approaches $0$ and is canceled by the corresponding pole, while the other tends toward infinity and thus has negligible impacts, such that we recover the ideal transferfunction $1$.
This is expected, since by setting the gains to zero the filter only integrates the angular velocity, resulting in the true angle.
However, the filter is not asymptotically stable, such that a constant bias in the gyroscope will also be integrated without any corrections.
As $k_p$ approaches infinity, the filter relies more on the accelerometer correction, and the zeros converge to $\pm\frac{1}{\sqrt{l\cos(\varphi_{\mathrm{op}})}}$, which matches the one-dimensional case in Section~\ref{sec:Analysis:OneDim}.
Note that computing $\arccos(\bar{a}^K_3)$ also yields $\varphi$, if the \ac{IMU} is at rest, but does not depend on $\ddot{\varphi}$.
Hence, the fact that the zeros align is a particular property of the $\mathrm{atan2}$.

\begin{figure}[tb]
	\centering
	\includegraphics[scale=0.45]{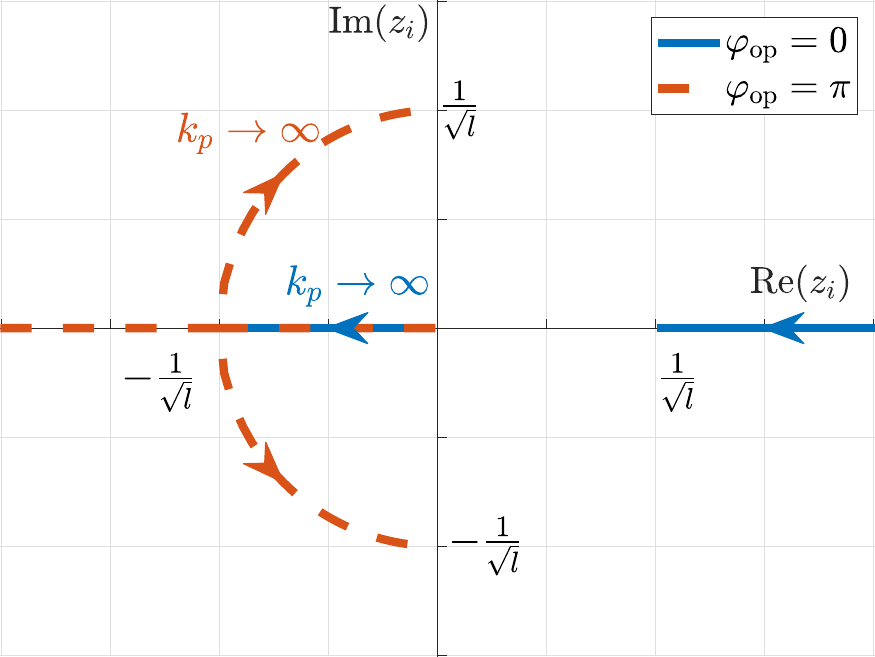}
	\caption{Location of the zeros 	for $\varphi_{\mathrm{op}}=0$ and $\varphi_{\mathrm{op}}=\pi$ for different $k_p$.}
	\label{IMG:Ana:ZerosMahony}
\end{figure}

For $\varphi_{\mathrm{op}}\in[0,\frac{\pi}{2})$, the system exhibits a qualitative change in the behavior of its zeros.
In particular, we get one zero with a negative real part and one with a positive real part, but both are shifted to the right.
Hence, the asymptotically stable zero is moved closer to the origin and already affects the estimation for lower velocities.
However, the unstable zero is farther away from the zero and hence, affects the estimation only for higher velocities.
In contrast, operating points $\varphi_{\mathrm{op}}\in(\frac{\pi}{2},\pi]$ change the qualitative behavior. 
The zeros are no longer on the imaginary axis, but are shifted to the left half plane.
Further, for $k_p \leq \frac{1}{\sqrt{2l\cos(\varphi_{\mathrm{op}})}}$, the zeros remain real-valued, and for $k_p > \frac{1}{\sqrt{2l\cos(\varphi_{\mathrm{op}})}}$ they get an imaginary part approaching $\pm i \frac{1}{\sqrt{l\cos(\varphi_{\mathrm{op}})}}$ as in the one-dimensional case.
A visualization on how the zeros are influenced by $k_p$ for $\varphi_{\mathrm{op}} = 0$ and $\varphi_{\mathrm{op}}=\pi$ is shown in Fig.~\ref{IMG:Ana:ZerosMahony}.
In both cases, $k_p$ must be chosen as a trade-off between fast poles for fast convergence and rejecting the effects of $\ddot{\varphi}$ by moving the zeros to a more desirable location.

\subsection{Madgwick filter}
In contrast to the Mahony filter, which is designed as a PI-controller, the Madgwick filter is formulated as an optimization problem to align the gravitational vector to the measured acceleration.
More precisely, the Madgwick filter uses a gradient descent algorithm with feedforward compensation for $\omega^K$ to minimize the cost function 
\begin{equation}
	f(\hat{q}, \bar{a}^K) = \left\|\hat{q}^\star \otimes \begin{pmatrix}0_{3\times1} \\ 1 \end{pmatrix} \otimes \hat{q} - \begin{pmatrix} 0 \\ \bar{a}^K\end{pmatrix}\right\|^2.
\end{equation}
This results in the following update rule for the estimated quaternion~\cite{Madgwick2011}
\begin{equation}\label{eq:ODE-Madgwick}
	\dot{\hat{q}} = \frac{1}{2} \hat{q} \otimes \begin{pmatrix}0 \\ \omega^K \end{pmatrix} -
	\beta \frac{\nabla f(\hat{q}, \bar{a}^K)}{\| \nabla f(\hat{q}, \bar{a}^K)\|}
\end{equation}
with step size $\beta$.
Unlike the Mahony filter, the differential equation~\eqref{eq:ODE-Madgwick} cannot be linearized directly due to the normalization of $\nabla f$, which introduces a discontinuity at the optimum $\nabla f = 0$. 
This non-smooth behavior is analogous to sliding mode observers, which operate on a sliding surface defined by $\nabla f=0$.
For the following analysis, we first assume $\beta$ to be sufficiently large to ensure that the filter tracks the sliding surface $\nabla f=0$ closely.
To analyze the behavior, we consider the sliding surface and linearize it around an operating point. 
Without loss of generality, we fix the yaw $\psi=0$ and the pitch $\theta=0$, which implies $q_{\mathrm{op},3}=q_{\mathrm{op},4}=0$ as operating points. 
Together with $\bar{a}^K_1=0$, this yields the linearized gradient
\begin{equation*}
	\nabla f \approx 
	\begin{pmatrix}
		4 & 0 &     0 &     0\\
		0 & 4 &     0 &     0\\
		0 & 0 & \star & \star\\
		0 & 0 & \star & \star
	\end{pmatrix}\Delta \hat{q} +
	\begin{pmatrix}
		-2 q_{\mathrm{op},2} & -2q_{\mathrm{op},1} \\
		-2 q_{\mathrm{op},1} & 2q_{\mathrm{op},2} \\
		                   0 & 0 \\
		                   0 & 0
	\end{pmatrix}
	\Delta \bar{a}^K_{2:3},
\end{equation*}
where $\star$ denotes components that are not relevant.
Assuming zero initial estimation error, we set $\Delta\hat{q}_3 = \Delta\hat{q}_4 = 0$, allowing us to solve explictly for $\Delta q_1$ and $\Delta q_2$ on the sliding surface $\nabla f = 0$.
Further, when linearizing the transformation from quaternion back to roll, pitch, and yaw around the operating point, it follows from $\frac{\partial \theta}{\partial q_1}=\frac{\partial \theta}{\partial q_2} = \frac{\partial \psi}{\partial q_1} = \frac{\partial \psi}{\partial q_2} = 0$ that the pitch and yaw are independent of the acceleration measurements. 
Hence, as for the Mahony filter, we deduce that violating the rest condition for one angle does not affect the estimation of the other angles.
This can be inserted into the linearized equation~\eqref{eq:analysis:phi}, yielding the estimated transfer function
\begin{equation}
	\Delta \hat{\varphi}(s) = (1-l\cos(\varphi_{\mathrm{op}})s^2) \Delta \varphi(s).
\end{equation}
This estimate is again equivalent to the one in~\eqref{eq:Ana:tfAtan} for the one-dimensional case. 
Now, we can apply the previous analysis to this sliding surface.
As with the Mahony filter, the Madgwick filter uses the gyroscope to mitigate the influence of $\ddot{\varphi}$.
For $\beta=0$, the filter integrates the gyroscope measurements without any error correction.
This results in an estimate insensitive to $\ddot{\varphi}$, but suffers from drift due to the gyroscope bias.
Due to the non-differentiable dynamics, a detailed analysis is as straightforward as for linear systems.
Due to the normalization, the filter can only move to the sliding surface with a rate of at most $\beta$.
If the sliding surface changes faster, the filter cannot track it accurately.
However, small $\beta$ generally leads to a filter which relies more on the integrated gyroscope values, and a larger $\beta$ yields an estimate that is closer to the sliding surface and may also compensate for the bias in the gyroscope measurements.
Hence, the step size $\beta$ acts as a tuning parameter to trade off fast convergence and rejection of $\ddot{\varphi}$.

\subsection{Discussion}
We emphasize once more that the difference between $\varphi$ and $\hat{\varphi}$ is not caused by an independent external disturbance.
For the linear case and by extension nonlinear case near to an operating point, such an external disturbance cannot alter system stability. 
However, $\ddot{\varphi}$ is not independent of $\varphi$.
This dependence introduces additional zeros into the transfer function, fundamentally altering the estimator's behavior for high frequencies.
This validates and explains previous observations, that the filters must be specifically tuned for each application to get an accurate estimate in the required frequency range \cite{Nazarahari2021}.

In feedback systems, the presence of zeros can have critical implications.
Especially, for unstable zeros, the controller design becomes more challenging and the achievable performance is limited due to the non-minimum phase behavior~\cite{Qiu1993,Freudenberg1985,Cheng1980,Misra1989}.
Asymptotically stable zeros, on the other hand, also degrade the estimate, but they introduce additional phase into the system, which can increase the phase margin and, hence, robustness.
Since unstable pole zero cancellation leads to a loss of internal stability, canceling undesirable zeros is only possible for the in the left-half plane.
Adding a slow pole for cancellation can help with the effects of $\ddot{\varphi}$, but decrease the overall bandwidth.
We emphasize that the location of zeros are independent of the controller, hence they can only be shifted by the estimator.

Our analysis suggests several possibilities to address these issues.
First, when designing the system, the \ac{IMU} should be placed close as possible to the rotational axis.
This reduces $l$ and moves the zeros to a higher frequency range to improve the behavior at intermediate frequencies.
Secondly, choose smaller values of  $k_p$ and $\beta$ to reduce sensitivity to the errors due to the accelerometer.
This comes at the cost of reduced bandwidth. 
Moreover, a model can be used to account for $\ddot{\varphi}$, e.g.,~\cite{Gajamohan2012} uses multiple \acp{IMU} with known kinematics to eliminate $\ddot{\varphi}$ from the measurements.
Finally, when designing a controller, the filter dynamics must be taken into account explictly to avoid loss of stability. 
Our analysis clearly shows that an independent design of estimator and controller, as allowed by the separation principle for linear systems, is not applicable in the considered setup.

\section{EXPERIMENT} \label{sec:Experiment}
In this section, we validate the theoretical results presented in the previous section.
To this end, we conduct an open-loop experiment with a dataset in Section~\ref{sec:Experiment-open-loop} and a closed-loop experiment with an autonomous \ac{e-scooter} in Section~\ref{sec:Experiment-closed-loop}.

\subsection{Open-Loop Experiment}\label{sec:Experiment-open-loop}

\begin{figure}[t]	
	\begin{minipage}{0.48\textwidth}
		\centering
		\includegraphics[scale=0.52]{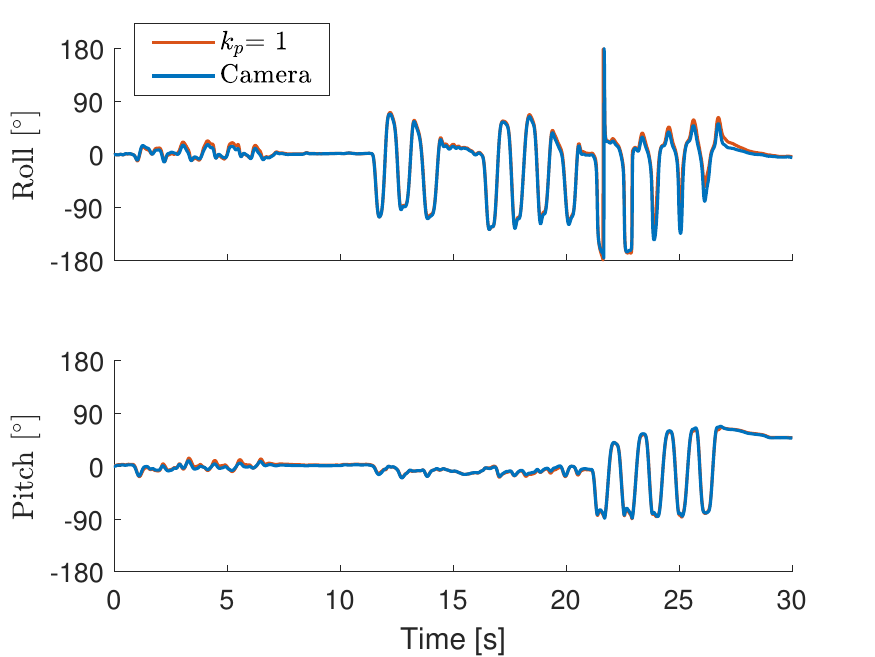}
		\caption{Sassari-Dataset with Mahony filter, $k_p = 1$.}
		\label{IMG:Exp:SasMahonyKp1}		
	\end{minipage}
	\begin{minipage}{0.04\textwidth}
	\end{minipage}
	\begin{minipage}{0.48\textwidth}
		\centering
		\includegraphics[scale=0.52]{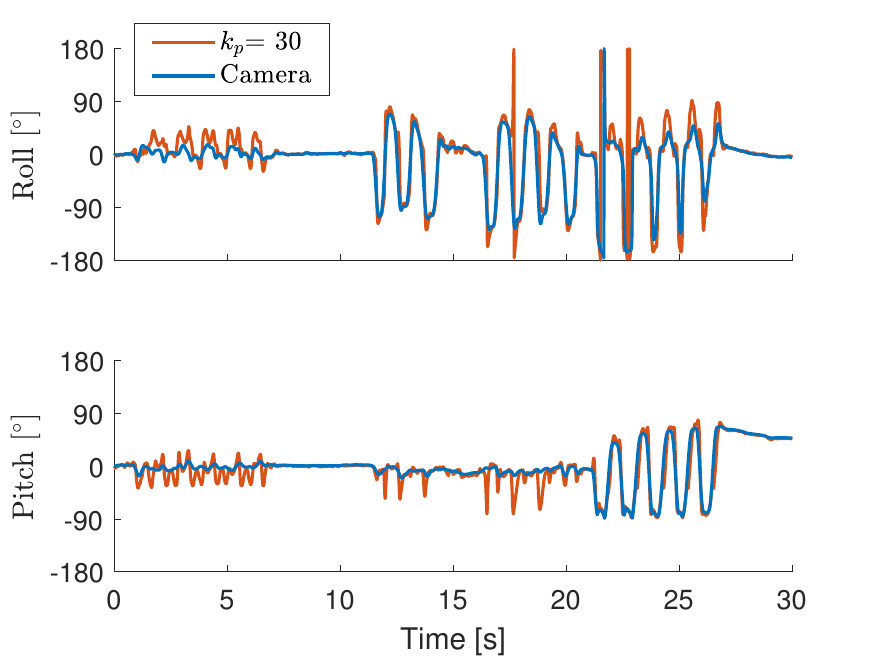}
		\caption{Sassari-Dataset with Mahony filter, $k_p = 30$.}
		\label{IMG:Exp:SasMahonyKp30}
	\end{minipage}		
\end{figure}

First, we consider the Sassari dataset~\cite{Caruso2021}, in which the \acp{IMU} are not placed at the rotational axes as in Section~\ref{sec:Analysis}. 
Among the available sensors, we focus on the first Xsens \ac{IMU}, since the others yield qualitatively similar results.
The dataset also provides orientation estimates from a camera system, which we treat as the ground truth, as it is unaffected by rotational accelerations.
Unlike to the previous sections, the dataset includes rotation around all axis.
Further, we only consider the \emph{fast} dataset with the largest angular accelerations.
As in the previous analysis, we do not consider the magnetometer to isolate the effects of the accelerometer and gyroscope interaction.
Fig.~\ref{IMG:Exp:SasMahonyKp1} and Fig.~\ref{IMG:Exp:SasMahonyKp30} show the simulation results for the Mahony-Filter with different $k_p$.
The Madgwick filter exhibits similar trends, and for brevity, we omit its plots.
Due to the large range of angles, providing a detailed analysis based on linearization is not possible, as the qualitative nature of the zeros is not the same for the whole range of motions.
As discussed in the previous section, choosing a small $k_p$ results in an estimator, which relies less on the accelerometer.
Hence, it attenuates the effects of the angular accelerations better.
Selecting a larger $k_p$ leads to a larger deviation between true orientation and estimate during fast motions.
This effect is particularly visible around $\SI{5}{\second}$, where the estimate significantly differs from the camera.

\subsection{Closed-Loop Experiment}\label{sec:Experiment-closed-loop}

\begin{figure}[t!]
	\centering
	\begin{minipage}{0.9\linewidth}
		\centering
		\hspace*{-0.1\linewidth}
		\input{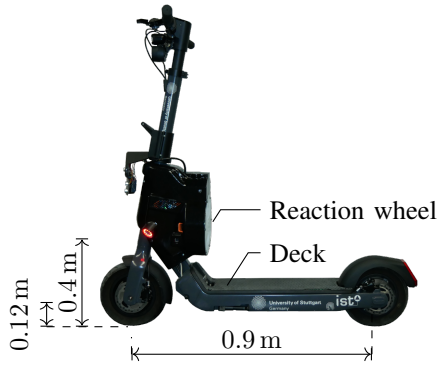}
	\end{minipage}
	\caption{Autonomous \ac{e-scooter}~\cite{Straesser2024}.}
	\label{IMG:Exp:EScooter}
\end{figure}

As a second example, we consider an autonomous \ac{e-scooter}~\cite{Soloperto2021,Straesser2025}, as illustrated in Fig.~\ref{IMG:Exp:EScooter}, to demonstrate the effects for feedback systems.
The \ac{e-scooter} is equipped with a reaction wheel to stabilize the upper equilibrium to enable autonomous driving.
To avoid falling over, the \ac{e-scooter} estimates the roll angle $\varphi$, which is fed into a cascaded PI- and PD-controller.
The controller calculates a desired motor current to simultaneously stop the \ac{e-scooter} from falling over and keep the reaction wheel speed low~\cite{Wenzelburger2020}.
The \ac{e-scooter} is controlled by two \emph{VESC 6 MK V} motor controllers.
Each \emph{VESC 6 MK V} has a built-in \emph{BMI 160} \ac{IMU} with a gyroscope and an accelerometer.
The first one is placed behind the reaction wheel at height $l = \SI{0.4}{\meter}$ and actuates the reaction wheel itself.
The second motor controller is located in the deck at height $l = \SI{0.12}{\meter}$ and ensures that the desired driving speed is maintained.
For comparison, the center of gravity is in between at a height of $\SI{0.17}{\meter}$.

\begin{figure*}[tb]	
	\begin{minipage}{0.48\textwidth}
		\centering
		\includegraphics[scale=0.53]{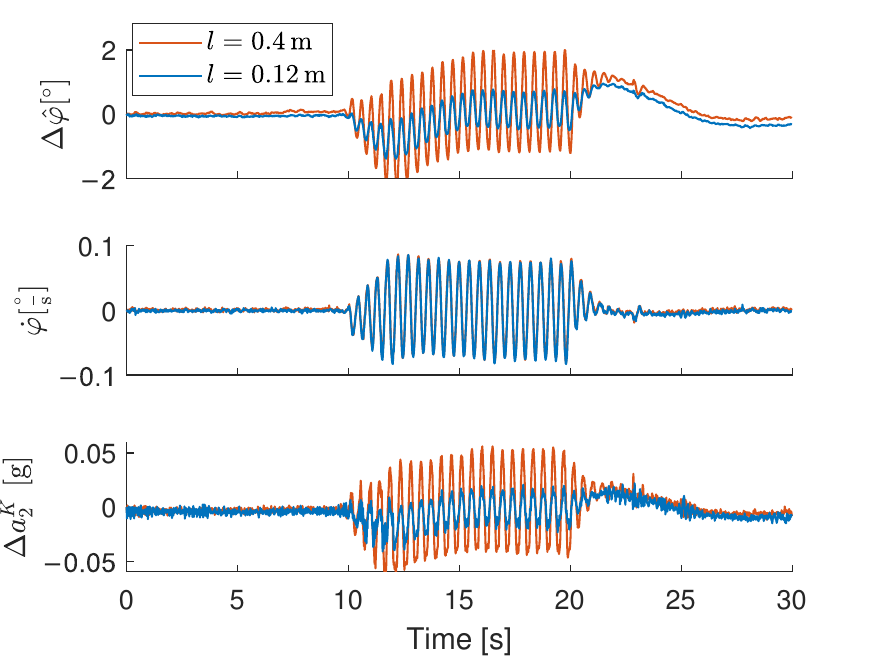}
		\caption{Closed loop with Mahony filter, $k_p = 10$, $k_i = 1$.}
		\label{IMG:Exp:MahonyKp10Ki1}		
	\end{minipage}
	\begin{minipage}{0.04\textwidth}
	\end{minipage}
	\begin{minipage}{0.48\textwidth}
		\centering
		\includegraphics[scale=0.53]{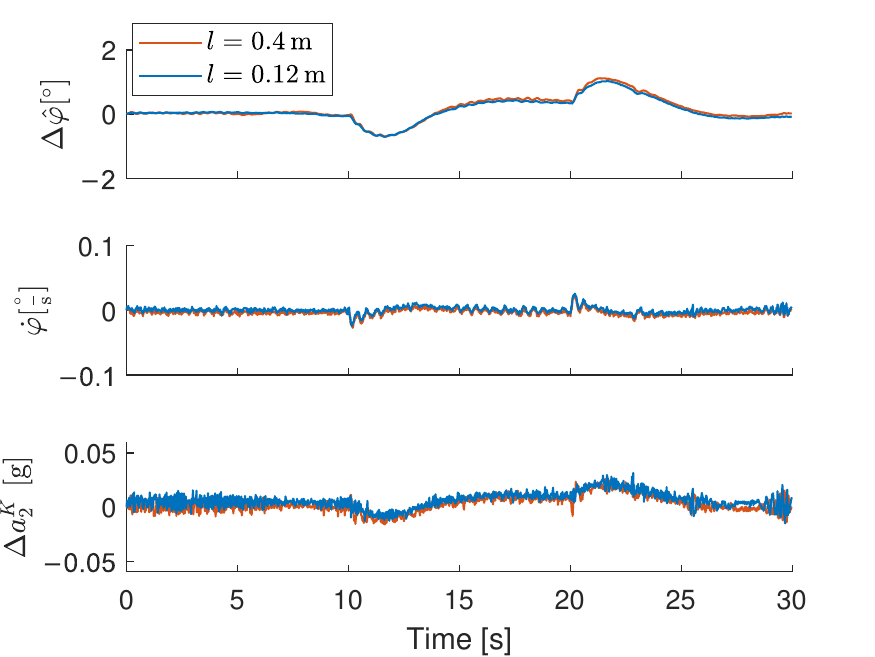}
		\caption{Closed loop with Mahony filter, $k_p = 2.2$, $k_i = 1$.}
		\label{IMG:Exp:MahonyKp2,2Ki1}
	\end{minipage}		
\end{figure*}

To investigate the effects of the \ac{IMU}, we let the scooter stabilize the upper equilibrium at standstill without any external excitation.
For the first $\SI{10}{\second}$ we use the estimated roll angle of the lower \ac{IMU}.
After $\SI{10}{\second}$ the controller switches to the estimate of the upper \ac{IMU} and switches back after another $\SI{10}{\second}$.
As estimation algorithm, we compare the Mahony filter for different values of $k_p$ and $k_i$.
Due to minor calibration mismatches in the accelerometer and a slight imbalance of the \ac{e-scooter}, the equilibrium is not exactly zero but is about $\pm 0.3^\circ$ depending on the \ac{IMU}.
For visualization purposes, we center the following plots around $0^\circ$ and indicate the mismatch by using $\Delta$ as the difference to the upper equilibrium.
Further, we only show $\bar{a}_2^K$, since this is the only measurement affected by $\ddot{\varphi}$.
Fig.~\ref{IMG:Exp:MahonyKp10Ki1} illustrates the resulting closed-loop behavior for $k_p = 10$ and $k_i = 1$.
For the first $\SI{10}{\second}$ the lower \ac{IMU} is used and the \ac{e-scooter} remains at the equilibrium.
However, when switching to the upper \ac{IMU}, the slight mismatch in the calibration of the accelerometers leads to an excitation of the system.
As indicated by the measurements of $\dot{\varphi}$, this results in fast oscillations in the roll angle, which do not vanish over time.
This oscillation also leads to significantly varying acceleration measurements $\Delta a^K_2$ and consequently to the varying estimated roll $\Delta \hat{\varphi}$.
As shown in Section~\ref{sec:Analysis}, the lower \ac{IMU} is less affected by this effect.
When switching back to the lower \ac{IMU} in the deck, the oscillations quickly vanish and the acceleration measurements and estimates $\Delta \hat{\varphi}$ resemble each other again.
This highlights the effect for feedback as unstable zeros are amplified in closed loop.
When the lower \ac{IMU} is used for control, the estimates of the upper and the lower \ac{IMU} behave similarly.
In this case, the upper \ac{IMU} operates in open loop and does not affect the closed loop. 
However, when using the upper \ac{IMU}, the non-minimum phase behavior is more prevalent and the estimation error is amplified leading to worse control performance.
To illustrate the effects of different filter parameters, we reduce $k_p$ to $k_p=2.2$ in Fig.~\ref{IMG:Exp:MahonyKp2,2Ki1}.
For this parameter set, there are no oscillations when employing the upper \ac{IMU} in the control loop.
The lower value of $k_p$ reduces the effect of the accelerometer, such that the filter relies more on the integrated measurements of the gyroscope and the corresponding zeros of the filter are farther away from the controller bandwidth.

Fig.~\ref{IMG:Exp:Madgwick_Beta0,1} and Fig.~\ref{IMG:Exp:Madgwick_Beta0,01} show the same experiment, but for the Madgwick filter with $\beta = 0.1$ and $\beta = 0.01$ instead.
The closed loop shows a similar behavior as for the Mahony filter.
Large values of $\beta$ increase the effects of $\ddot{\varphi}$ and result in fast oscillations when the upper \ac{IMU} is employed in the closed loop.
A smaller $\beta$ attenuates the angular acceleration by relying more on the gyroscope and less on the accelerometer, but requires a sufficiently accurate gyroscope.

\begin{figure*}[tb]	
	\begin{minipage}{0.48\textwidth}
		\centering
		\includegraphics[scale=0.53]{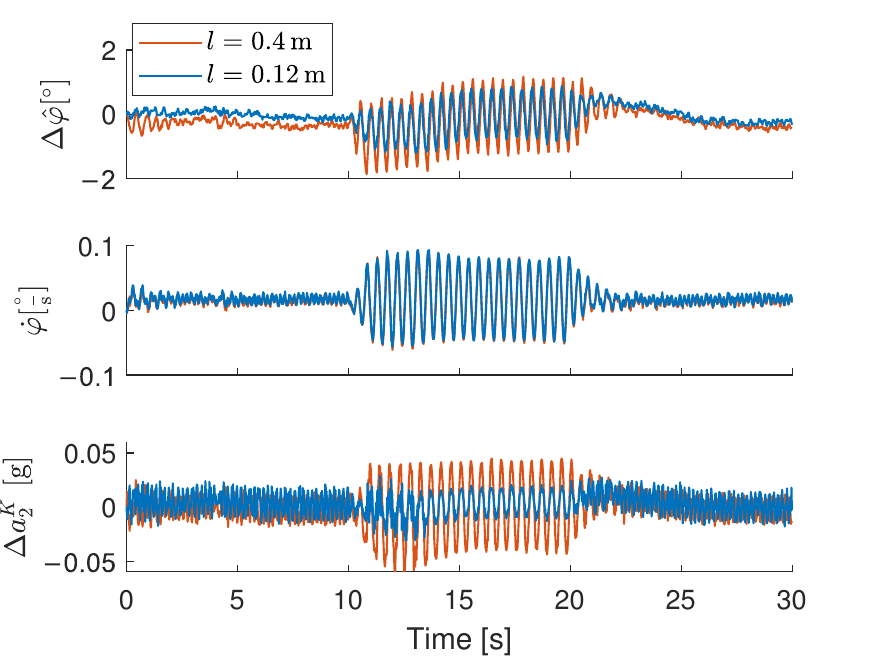}
		\caption{Closed loop with Madgwick filter, $\beta = 0.1$.}
		\label{IMG:Exp:Madgwick_Beta0,1}		
	\end{minipage}
	\begin{minipage}{0.04\textwidth}
	\end{minipage}
	\begin{minipage}{0.48\textwidth}
		\centering
		\includegraphics[scale=0.53]{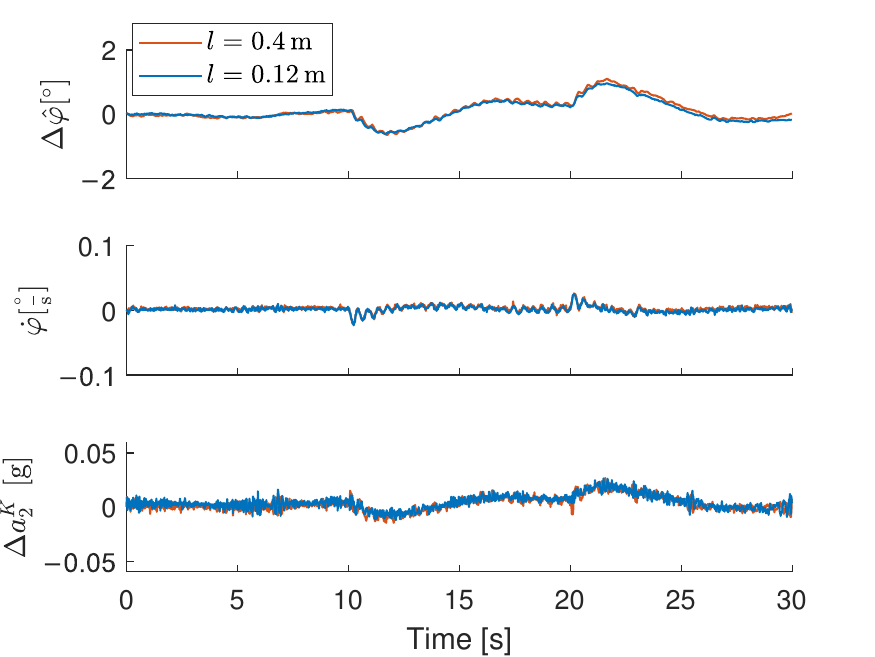}
		\caption{Closed loop with Madgwick filter, $\beta = 0.01$.}
		\label{IMG:Exp:Madgwick_Beta0,01}
	\end{minipage}		
\end{figure*}

With the conducted experiments, we are able to recover the theoretical results of Section~\ref{sec:Analysis}.
To further reduce the effects of $\ddot{\varphi}$, multiple \acp{IMU} can be used~\cite{Gajamohan2012}.
Alternatively, if there are no further sensors available and the existing sensor cannot be placed closer to the rotational axis, one possibility is to include the estimator in the controller synthesis.
Treating the angular accelerations as external disturbances does not capture its true effects.
Hence, the controller design must explicitly account for it in the model or the the designed controller must be robust against this model uncertainty.
Further, we want stress to again, an estimator which shows good results in open-loop may not be suited to be used in closed-loop with a controller.
This also requires a new way to validate filters.
As datasets with pre-collected data can not accurately represent the effects of angular accelerations, such that an estimator performing well in open loop may show low performance in closed loop.

\section{OUTLOOK} \label{sec:Outlook}
In this work, we investigated rotational accelerations and their effect on different sensor fusion algorithms based on \acp{IMU}.
We showed that this can lead to non-minimum phase behavior, which is known to be challenging in feedback systems. 
Further, we analyzed how common filter methods behave and how appropriate parameter tuning can mitigate these effects.
However, mitigation comes at the cost of a reduced filter bandwidth.
Finally, we validated our theoretical findings on a real system by showing that the performance in closed loop can significantly deviate from the open loop.
Hence, the controller and estimator cannot be designed independently.
Future work should look into designing new benchmarks to accurately capture closed loop effects.

\bibliographystyle{IEEEtran}
\bibliography{CCTAScooterIMU}

\end{document}